\documentclass[12pt]{article}
\usepackage{a4wide,epsfig,psfrag,amsmath,amssymb,cite,scalefnt}
\usepackage{color}

\graphicspath{ {figs/} }

\newcommand{\gsim}{\;\rlap{\lower 3.5 pt \hbox{$\mathchar \sim$}} \raise 1pt
 \hbox {$>$}\;}
\newcommand{\lsim}{\;\rlap{\lower 3.5 pt \hbox{$\mathchar \sim$}} \raise 1pt
 \hbox {$<$}\;}

\allowdisplaybreaks

\begin{document}

\title{\vskip-3cm{\baselineskip14pt
    \begin{flushleft}
      \normalsize SFB/CPP-13-81\\
      \normalsize TTP13-32 \\
      \normalsize DESY 13-195 \\
      \normalsize LPN13-084 \\
  \end{flushleft}}
  \vskip1.5cm
  Anomalous magnetic moment with heavy virtual leptons
}

\author{
  Alexander Kurz$^{a,b}$,
  Tao Liu$^{a}$,
  Peter Marquard$^{b}$,
  Matthias Steinhauser$^{a}$
  \\[1em]
  {\small\it (a) Institut f{\"u}r Theoretische Teilchenphysik,}
  {\small\it Karlsruhe Institute of Technology (KIT)}\\
  {\small\it 76128 Karlsruhe, Germany}
  \\
  {\small\it (b) Deutsches Elektronen Synchrotron (DESY),}\\
  {\small\it 15738 Zeuthen, Germany}
}

\date{}

\maketitle

\thispagestyle{empty}

\begin{abstract}

  We compute the contributions to the electron and muon anomalous magnetic
  moment induced by heavy leptons up to four-loop order. Asymptotic expansion
  is applied to obtain three analytic expansion terms which show rapid
  convergence.

  \medskip

\end{abstract}

\thispagestyle{empty}




\section{Introduction}

Since decades the anomalous magnetic moments of electron and muon, $a_e$
and $a_\mu$, are used to perform precision tests of QED.\footnote{ See
  Refs.~\cite{Melnikov:2006sr,Jegerlehner:2009ry,Miller:2012opa} for
  comprehensive reviews.}  In fact, in the case of the electron the
experimental measurements and theoretical predictions have reached a precision
which allows for the most precise extraction of the fine structure constant
$\alpha$. In contrast to $a_e$ there is a sizable hadronic contribution to
$a_\mu$ which involves as input measurements of the total cross section
$\sigma(e^+e^-\to\mbox{hadrons})$ at low energies.  Although all ingredients
are measured and computed to high precision there is a discrepancy of about
$3\sigma$ between the measured and predicted value for
$a_\mu$~\cite{Jegerlehner:2011ti,Hagiwara:2011af,Davier:2010nc}.  In this
context it is interesting to mention that this difference is of the same order
of magnitude as the four-loop QED contribution which to date has only been
computed by one group~\cite{Aoyama:2012wk}. Thus, it is important to provide
an independent cross check for this ingredient.  First results have been
obtained in Ref.~\cite{Laporta:1993ds,Aguilar:2008qj,Lee:2013sx}. In
particular, in Ref.~\cite{Lee:2013sx} the contribution from Feynman diagrams
containing two or three closed electron loops have been computed.  In this
letter we provide a further step towards the full four-loop QED corrections to
$a_e$ and $a_\mu$ and compute the part induced by heavy leptons.  In the case
of the muon this means that Feynman diagrams have to be considered which
contain closed tau loops and both closed muon and tau loops are present 
for  $a_e$.
Such contributions appear for the first time at two-loop order
(cf. Fig.~\ref{fig::diags}) and have been computed in
Refs.~\cite{Elend:1966,Passera:2004bj,Passera:2006gc}.  Also the three-loop
result is known in analytic form for arbitrary lepton
masses~\cite{Samuel:1990qf,Laporta:1992pa,Laporta:1993ju,Czarnecki:1998rc,Kuhn:2003pu,Friot:2005cu}
(see also~\cite{Passera:2004bj,Passera:2006gc}). At four loops, however, only
numerical results are
available~\cite{Kinoshita:2005zr,Aoyama:2007mn,Aoyama:2012wj,Aoyama:2012wk}.
We want to cross-check these results using
a different method which leads to analytic results
for $a_e$ and $a_\mu$.  It is based on asymptotic
expansion~\cite{Smirnov:2013} in the ratio of the light and heavy lepton mass,
$M_l$ and $M_h$, which leads to a factorization of the two-scale integrals
into simpler ones with at most one mass scale.  The latter can be computed
analytically.  We have computed three terms of the expansion in $M_l^2/M_h^2$.

For the perturbative expansion of the QED corrections to $a_e$ and $a_\mu$ we
take over the commonly used notation from
Refs.~\cite{Aoyama:2012wj,Aoyama:2012wk} and write ($l=e,\mu$)
\begin{eqnarray}
  a_l &=& \sum_{n\ge1} \left(\frac{\alpha}{\pi}\right)^n a_l^{(2n)}
  \,,
\end{eqnarray}
where $a_l^{(2n)}$ can be written in the form
\begin{eqnarray}  
  a_e^{(2n)} &=&  A_{1,e}^{(2n)} 
  + A_{2,e}^{(2n)}(M_e/M_\mu) 
  + A_{2,e}^{(2n)}(M_e/M_\tau) 
  + A_{3,e}^{(2n)}(M_e/M_\mu,M_e/M_\tau) 
  \,,\nonumber\\
  a_\mu^{(2n)} &=&  A_{1,\mu}^{(2n)} 
  + A_{2,\mu}^{(2n)}(M_\mu/M_e) 
  + A_{2,\mu}^{(2n)}(M_\mu/M_\tau) 
  + A_{3,\mu}^{(2n)}(M_\mu/M_e,M_\mu/M_\tau) 
  \,.
\end{eqnarray}
In this paper we compute $A_{2,e}^{(2n)}(M_e/M_\mu)$,
$A_{2,e}^{(2n)}(M_e/M_\tau)$, $A_{3,e}^{(2n)}(M_e/M_\mu,M_e/M_\tau)$ and
$A_{2,\mu}^{(2n)}(M_\mu/M_\tau)$ to four-loop order.
We have also computed the corresponding two- and three-loop results and found
complete agreement with the literature.

Before starting the actual calculation let us consider the parametrical size
of our corrections. Actually, the heavy-lepton contribution decouples in
the limit $M_h\to\infty$ and leads to a $M_l^2/M_h^2$ suppression. Thus the
four-loop corrections to $a_\mu$ have the form $(\alpha/\pi)^4 \times
M_\mu^2/M_\tau^2$ where $M_\mu^2/M_\tau^2$ is of order $10^{-3}$. On the other
hand we have $\alpha/\pi \approx 2\cdot 10^{-3}$ which is of the same order of
magnitude. Thus, from the parametric point of view the four-loop
corrections induced by
heavy leptons could be of the same order as the five-loop results obtained in
Ref.~\cite{Aoyama:2012wk}.  Note, however, that in practice the contributions
involving electron loops are large ($A_{2,e}^{(10)}(M_e/M_\mu)$ is of order
$10^3$)  whereas the heavy-lepton
contributions have coefficients which are at most of order $10$.

In the case of $a_e$ the ratio of the lepton masses is much smaller than for
$a_\mu$ (Note that $M_e^2/M_\mu^2 = {\cal O}(10^{-5})$, $M_e^2/M_\tau^2 = {\cal
  O}(10^{-8})$) and thus the corresponding corrections are less
relevant. Nevertheless, for completeness we provide also those results.

The remainder of the paper is structured as follows: In the next Section we
briefly discuss some technical details which are important for our calculation.
Section~\ref{sec::res} is devoted to the presentation and discussion of the
results. In particular, we compare to the numerical results of
Refs.~\cite{Aoyama:2012wj,Aoyama:2012wk}. We conclude in
Section~\ref{sec::concl}.
In the Appendix we present results for the on-shell counterterms
for the fine structure constant, the lepton mass and the lepton wave
function. 


\section{Some technical details}

\begin{figure}[t]
  \begin{center}
      \leavevmode
      \epsfxsize=0.9\textwidth
      \epsffile[80 390 500 750]{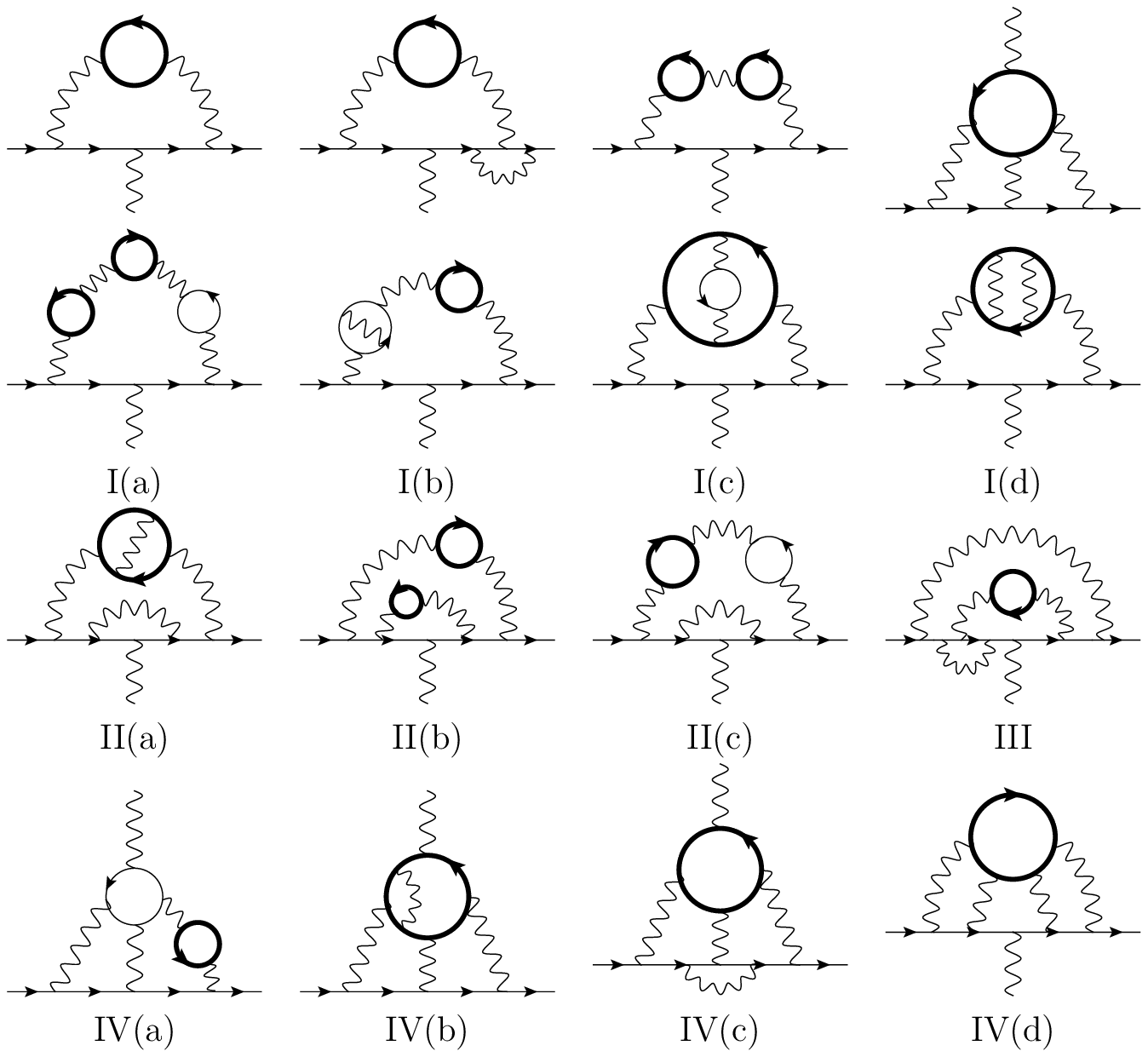}
  \end{center}
  \caption[]{\label{fig::diags} Sample Feynman diagrams contributing to the
    electron and muon $(g-2)$ containing a heavy lepton at two, three and four
    loops. Thin and thick solid lines represent light and heavy leptons,
    respectively, and wavy lines denote photons. The symbols below the
    four-loop diagrams label the individual diagram classes and are taken over
    from Refs.~\cite{Aoyama:2012wj,Aoyama:2012wk}.}
\end{figure}

Typical Feynman diagrams to be considered for the heavy-lepton contribution of
$a_e$ and $a_\mu$ are shown in Fig.~\ref{fig::diags}. At two-loop order only
one diagram has to be considered.\footnote{Note that the contribution
  where the photon couples to the closed lepton loop vanishes due to Furry's
  theorem~\cite{Furry:1937zz}.} At three loop-order 60 and at four loops 1169
Feynman diagrams are generated. In the following discussion we denote the
heavy lepton mass by $M_h$ and the light one by $M_l$.

For the generation of the diagrams we use {\tt QGRAF}~\cite{Nogueira:1991ex}
and transform the amplitudes with the help of {\tt
  q2e}~\cite{Harlander:1997zb,Seidensticker:1999bb} to a {\tt
  FORM}~\cite{Vermaseren:2000nd} readable output.

\begin{figure}[t]
  \begin{center}
    \begin{tabular}{c}
      \leavevmode
      \epsfxsize=0.9\textwidth
      \epsffile[80 600 500 750]{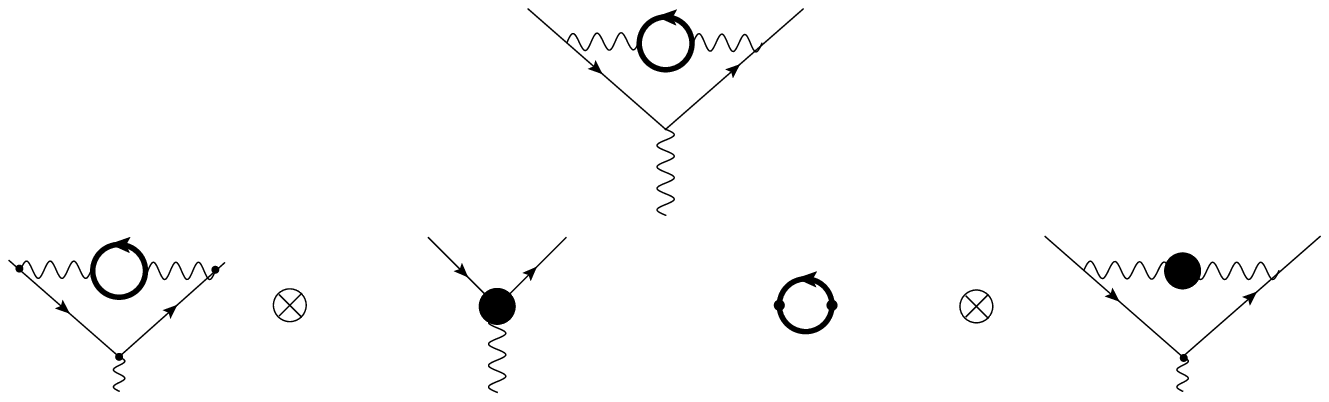}
      \\
      (a)
      \\
      \leavevmode
      \epsfxsize=0.9\textwidth
      \epsffile[80 500 550 750]{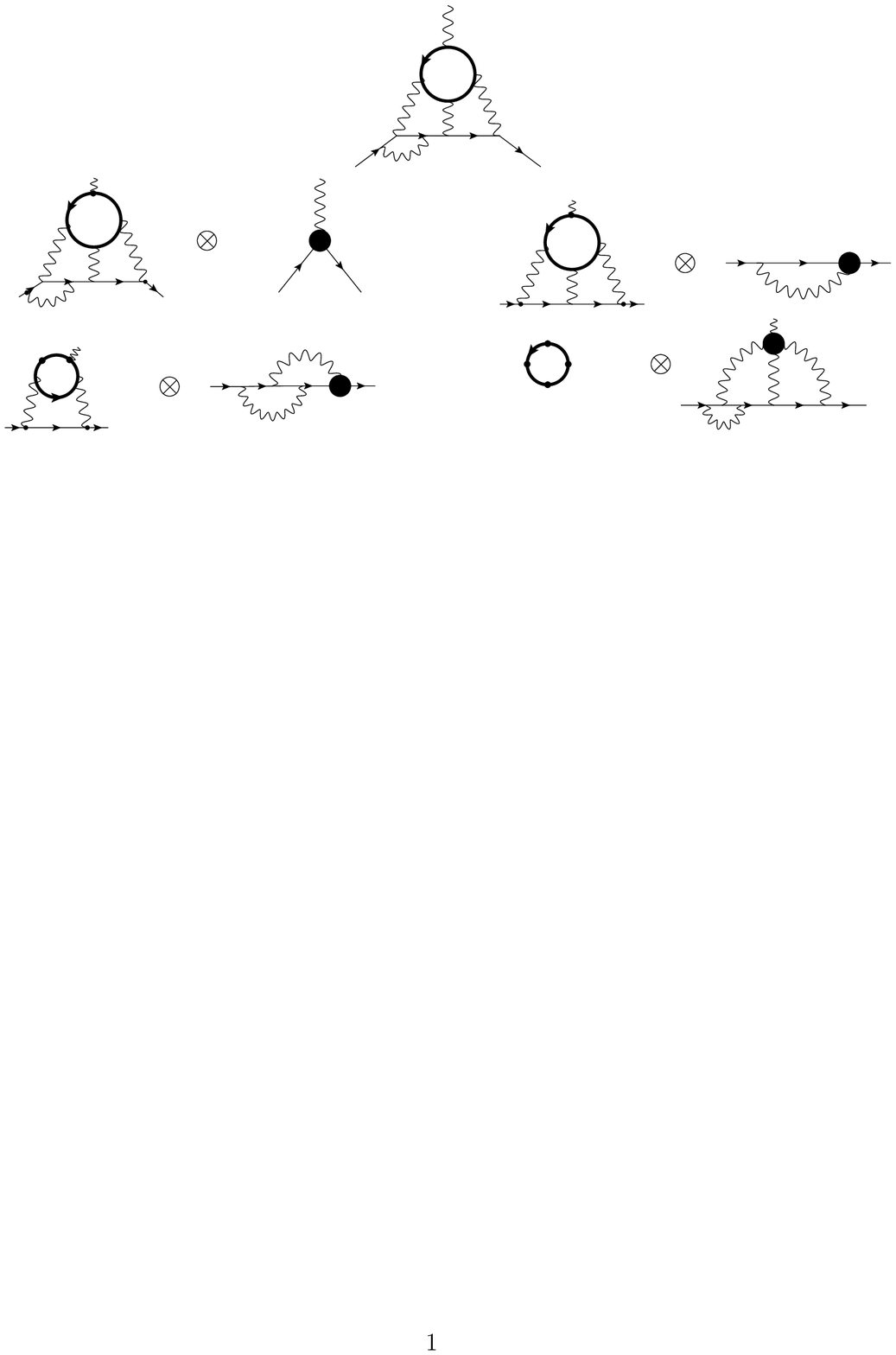}
      \\
      (b)
    \end{tabular}
  \end{center}
  \caption[]{\label{fig::ae} Graphical examples for the application of the
    asymptotic expansion at two (a) and four (b) loops. Thick solid, thin
    solid and wavy lines represent heavy and light leptons and photons,
    respectively. In (b) only four representative sub-diagrams are
      shown. Altogether there are eight contributions.} 
\end{figure}

In a next step we apply {\tt exp}~\cite{Harlander:1997zb,Seidensticker:1999bb}
to perform an asymptotic expansion for $M_h\gg M_l$.  At two-loop order [see
Fig.~\ref{fig::ae}(a)] this leads to two so-called sub-diagrams which have to
be Taylor-expanded in their external momenta. The first sub-diagram is given
by the whole two-loop diagram which, after expansion, leads to two-loop vacuum
integrals. The second contribution consists of a product of two one-loop
diagrams.  After expanding the one-loop vacuum integral in the external
momentum one has to insert the result in the remaining one-loop on-shell
integral and integrate over the second loop momentum.  The described procedure
is illustrated in the second line of
Fig.~\ref{fig::ae}(a).  Fig.~\ref{fig::ae}(b) shows a
four-loop example which demonstrates the typical situation at this order: the
original four-loop two-scale integral is transformed to a sum of products of
$N$-loop vacuum integrals with scale $M_h$ and $(4-N)$-loop on-shell integrals
with $q^2=M_l^2$ where $N=1,2,3$ or $4$ and $q$ is the momentum flowing
through the external lepton line.  All integrals only contain one mass scale
and are thus significantly simpler than the original one.

Both vacuum and on-shell integrals are reduced to master integrals
with the help of {\tt FIRE}~\cite{Smirnov:2008iw,Smirnov:2013dia}.\footnote{We
  thank A.V.~Smirnov and V.A.~Smirnov for allowing us to use the unpublished
  {\tt C++} version of {\tt FIRE}.} The master integrals are all known
analytically and are taken from Refs.
\cite{Laporta:2002pg,Chetyrkin:2004fq
,Kniehl:2005yc
,Schroder:2005db
,Schroder:2005hy
,Schroder:2005va
,Bejdakic:2006vg
,Chetyrkin:2006dh
,Kniehl:2006bf
,Kniehl:2006bg} and 
Refs.~\cite{Laporta:1996mq,Melnikov:2000qh,Lee:2010ik}, respectively. 

We renormalize our results in the on-shell scheme. For this purpose we need
the counter\-term for the fine structure constant, the (light)
lepton mass and lepton
wave function to three loops.  The corresponding analytic results for the
  case of a massless second lepton loop can be found in
  Ref.~\cite{Broadhurst:1991fi},
  Refs.~\cite{Chetyrkin:1999ys,Chetyrkin:1999qi,Melnikov:2000qh,Marquard:2007uj}
  and~\cite{Melnikov:2000zc,Marquard:2007uj}, respectively. In our case the
opposite limit of a heavy lepton is needed which we computed ourselves using
the rules of asymptotic expansion as described above.  The analytic
expressions are presented in the Appendix for completeness.  Our results for
the leading term of the lepton mass counterterm agrees with
Ref.~\cite{Sturm:2013uka} and the one for the charge counterterm is easily
obtained from the general expression presented in Ref.~\cite{Lee:2013sx}. To
our knowledge the three-loop result for the on-shell wave function
renormalization constant is new.

In addition, the heavy-lepton mass has to be renormalized in the two- and
three-loop expression. The corresponding two-loop counterterm can be found in
Refs.~\cite{Gray:1990yh,Bekavac:2007tk}.  Note that the two-loop counterterm
which has to be inserted into the two-loop vertex diagram of
Fig.~\ref{fig::diags} involves contributions with a closed light lepton
loop. The expansion of this contribution in $M_l\ll M_h$ contains both even
and odd powers in $M_l/M_h$ which is the reason for the occurrence of odd
expansion terms in $A^{(8)}_{2,\mu}$ (cf. Section~\ref{sec::res}).

There are several checks on the correctness of our result. Besides the obvious
ones like finiteness we have performed two independent calculations. In
particular, two independent routines for the decomposition of the scalar
products in the numerator and the preparation of the {\tt FIRE} input has been
written.  Furthermore for our calculation we have used general QED gauge
parameter up to linear terms in $\xi$ and have checked that the final result of the
leading term in the inverse heavy lepton expansion, i.e. the one proportional
to $M_l^2/M_h^2$, is $\xi$-independent. Due
to the complexity of the calculation we have used Feynman gauge for the higher
order expansion terms.


\section{\label{sec::res}Results and discussion}

Let us in a first step present the analytic results of our calculation.
The four-loop contribution to $a_\mu$ from Feynman diagrams involving a
virtual tau lepton loop is given by
\begin{eqnarray}
  A^{(8)}_{2,\mu}(M_\mu/M_\tau) &=&
  \left(\frac{M_\mu}{M_\tau}\right)^2 \bigg(\frac{37448693521}{2286144000}+\frac{89603}{16200}P_4
  +\frac{52}{675}P_5+\frac{4 \pi^2 \zeta_3}{15}+\frac{5771 \ln(2) \pi^4}{32400}\nonumber\\&&
  \qquad-\frac{3851 \pi^2}{3600}-\frac{25307 \zeta_5}{1440}-\frac{37600399 \pi^4}{27216000}+\frac{35590996657 \zeta_3}{508032000}\nonumber\\&&
  \qquad+\ln\frac{M_\mu^2}{M_\tau^2} \left(-\frac{38891}{12150}+\frac{19 \pi^2}{135}+\frac{3 \zeta_3}{2}\right)+\frac{359}{1080}\ln^2\frac{M_\mu^2}{M_\tau^2}\bigg)\nonumber\\&&
  + \left(\frac{M_\mu}{M_\tau}\right)^3 \frac{\pi^2}{90}\nonumber\\&&
  + \left(\frac{M_\mu}{M_\tau}\right)^4 \bigg(\frac{392783023945426851403}{73077446697615360000}-\frac{3355249339331 \pi^4}{2575112601600}\nonumber\\&&
  \qquad+\frac{74184592369}{14306181120} P_4
  +\frac{557}{9450}P_5\nonumber\\&&
  \qquad-\frac{378681587 \pi^2}{114307200}-\frac{652 \ln(2) \pi^2}{1215}+\frac{26783 \ln(2) \pi^4}{226800}\nonumber\\&&
  \qquad+\frac{725750082915523417 \zeta_3}{10310750856806400}+\frac{66211 \pi^2 \zeta_3}{22680}-\frac{425983 \zeta_5}{30240}\nonumber\\&&
  \qquad+\ln\frac{M_\mu^2}{M_\tau^2} \left(-\frac{1922512966823}{1229031014400}+\frac{47899 \pi^2}{816480}+\frac{81782993 \zeta_3}{123863040}\right)\nonumber\\&&
  \qquad+\frac{193032971}{457228800}\ln^2\frac{M_\mu^2}{M_\tau^2}-\frac{24037}{362880}\ln^3\frac{M_\mu^2}{M_\tau^2}\bigg)\nonumber\\&&
  + \left(\frac{M_\mu}{M_\tau}\right)^5 \left(\frac{2671 \pi^2}{176400}+\frac{\pi^2}{140} \ln\frac{M_\mu^2}{M_\tau^2}\right)\nonumber\\&&
  + \left(\frac{M_\mu}{M_\tau}\right)^6 \bigg( \frac{326292200455466311953239}{4974581098834034688000}+\frac{4785889811617 \pi^2}{1234517760000}\nonumber\\&&
  \qquad+\frac{989648650006997}{191294078976000} P_4
+\frac{7001}{207900}P_5\nonumber\\&&
  \qquad-\frac{27903657664078117 \pi^4}{11477644738560000}+\frac{711883 \ln(2) \pi^4}{9979200}-\frac{148 \ln(2) \pi^2}{315}\nonumber\\&&
  \qquad+\frac{6446695611351419899 \zeta_3}{66315280711680000}-\frac{18533 \pi^2 \zeta_3}{6048}+\frac{179971 \zeta_5}{24192}\nonumber\\&&
  \qquad+\ln\frac{M_\mu^2}{M_\tau^2} \bigg(-\frac{2631561259843654279}{132735349555200000}+\frac{17955349 \pi^2}{489888000}\nonumber\\&&
  \qquad\qquad\qquad+\frac{314284167899 \zeta_3}{19818086400}\bigg)\nonumber\\&&
  \qquad+\frac{22710352067}{58786560000} \ln^2\frac{M_\mu^2}{M_\tau^2}-\frac{101799017}{979776000} \ln^3\frac{M_\mu^2}{M_\tau^2}\bigg)\nonumber\\&&
  + \left(\frac{M_\mu}{M_\tau}\right)^7 \left(\frac{79 \pi^2}{15120}+\frac{\pi^2}{60} \ln\frac{M_\mu^2}{M_\tau^2}\right)
  + {\cal O}\left( \left(\frac{M_\mu}{M_\tau}\right)^8  \right)
  \nonumber \\
  &\approx&
  0.0421670 + 0.0003257 + 0.0000015
  \,,
  \label{eq::A8}
\end{eqnarray}
where $P_4=24a_4+\ln^4(2)-\ln^2(2) \pi^2$,
$P_5=120a_5-\ln^5(2)+\frac{5}{3}\ln^3(2) \pi^2$, $a_n=\mbox{Li}_n(1/2)$ and $\zeta_n$ is Riemann's zeta function.
In the last line of Eq.~(\ref{eq::A8}) the analytic expression has been
evaluated numerically using 
\mbox{$M_\mu/M_\tau=5.94649(54) \cdot 10^{-2}$}~\cite{Beringer:1900zz}.
Furthermore the
contributions from 
$(M_\mu/M_\tau)^n$ and $(M_\mu/M_\tau)^{n+1}$ ($n=2,4,6$) have been
combined. One observes a rapid convergence of the series in $M_\mu/M_\tau$
which suggests that with each additional order one gains two significant
digits. To be conservative we take 10\% of the last term in Eq.~(\ref{eq::A8})
as error estimate which leads to our final result
\begin{eqnarray}
  A^{(8)}_{2,\mu}(M_\mu/M_\tau)&\approx&
  0.0424941(2)(53)
  \,,
  \label{eq::A8mu}
\end{eqnarray}
where the second uncertainty reflects the error in the input quantity
$M_\mu/M_\tau$. 
The result in~(\ref{eq::A8mu}) agrees with the one from
Ref.~\cite{Aoyama:2012wk} $A^{(8)}_{2,\mu}(M_\mu/M_\tau) = 0.04234(12)$,
however, our number is significantly more precise. 

For completeness we also provide the numerical results for the two- and
three-loop contributions which read\footnote{Since analytic expressions are
  available the uncertainties for the two- and three-loop results are due to
  the errors in the lepton masses.}
\begin{eqnarray}
  A^{(4)}_{2,\mu}(M_\mu/M_\tau) &=& 7.8079(14) \cdot 10^{-5}
  \,,\nonumber\\
  A^{(6)}_{2,\mu}(M_\mu/M_\tau) &=& 3.6063(12) \cdot 10^{-4}
  \,.
\end{eqnarray}
It is interesting to note that the three-loop coefficient is only a factor
of five larger than the two-loop one whereas $A^{(8)}_{2,\mu}(M_\mu/M_\tau)$
is about 100 times larger than $A^{(6)}_{2,\mu}(M_\mu/M_\tau)$.
Using $\alpha = 1/137.035999174$~\cite{Aoyama:2012wj}
one finally obtains for the $\tau$-loop contribution to $a_\mu$
\begin{eqnarray}
  10^{11} \times a_\mu\Big|_{\tau \rm loops}
  &=& 42.13 + 0.45 + 0.12
  \,,
  \label{eq::amu_tau}
\end{eqnarray}
where the numbers on the right-hand side correspond to the two-, three- and
four-loop contribution.
The numbers in Eq.~(\ref{eq::amu_tau}) have to be compared with the 
universal contributions contained in $A_{1,\mu}$ which
read~\cite{Aoyama:2012wk}
\begin{eqnarray}
  10^{11} \times a_\mu\Big|_{\rm univ.} \!\!\!&=&\!\!\! 
  116\,140\,973.21 - 177\,230.51 + 1\,480.42 - 5.56 + 0.06
  \,,
\end{eqnarray}
where the individual terms on the right-hand side 
represent the results from one to five loops.

\begin{table}[t]
  \begin{center}
    \begin{tabular}{c||l|l}
      group & \multicolumn{2}{c}{$10^2 \cdot A^{(8)}_{2,\mu}(M_\mu/M_\tau)$}\\
      \hline
      & this work & \cite{Aoyama:2012wk}\\
      \hline
      I(a) & \hphantom{$-$}0.00324281(2) & \hphantom{$-$}0.0032(0)\\
      I(b) + I(c) + II(b) + II(c) & $-0.6292808(6)$ & $-0.6293(1)$\\
      I(d) & \hphantom{$-$}0.0367796(4) & \hphantom{$-$}0.0368(0)\\
      III & \hphantom{$-$}4.5208986(6) & \hphantom{$-$}4.504(14)\\
      II(a) + IV(d) & $-2.316756(5)$ & $-2.3197(37)$\\
      IV(a) & \hphantom{$-$}3.851967(3) & \hphantom{$-$}3.8513(11)\\
      IV(b) & \hphantom{$-$}0.612661(5) & \hphantom{$-$}0.6106(31)\\
      IV(c) & $-1.83010(1)$ & $-1.823(11)$
    \end{tabular}
  \end{center}
  \caption{\label{tab::mu}Mass-dependent corrections to
    $a_\mu$ at four-loop order as obtained in this paper and the 
    comparison with Refs.~\cite{Aoyama:2012wk}.
    The uncertainties assigned to our numbers correspond to 
    10\% of the highest available expansion terms, i.e., the 
    ones of order $(M_\mu/M_\tau)^6$ and $(M_\mu/M_\tau)^7$.
    Uncertainties from the muon and tau lepton mass are not shown.}
\end{table}

The detailed comparison with Tab.~I of Ref.~\cite{Aoyama:2012wk} is shown in
Tab.~\ref{tab::mu} where our result is split into eight different groups.
In the first column the notation of~\cite{Aoyama:2012wk} is used to indicate 
the contributions which have to be summed\footnote{We add the uncertainties of
  Ref.~\cite{Aoyama:2012wk} in quadrature when adding results from different
  groups.} in order to compare with our 
numbers. Within the numerical uncertainties we observe good agreement.
Note, however, that our results based on asymptotic expansion provide at least
two more significant digits. 

Let us mention that the analytic result for the 
leading order expansion term of case IV(b) agrees with the result presented in
Ref.~\cite{Kataev:2012kn} which has been obtained by transforming the 
result of Ref.~\cite{Boughezal:2011vw} to QED.

Let us next turn to the anomalous magnetic moment of the electron.
The numerical values for the two- and three-loop contributions read
  \begin{eqnarray}
    A^{(4)}_{2,e}(M_e/M_\mu) &=& 5.19738668(26) \cdot 10^{-7}
    \,,\nonumber\\
    A^{(6)}_{2,e}(M_e/M_\mu) &=& - 7.37394162(27) \cdot 10^{-6}
    \,,\nonumber\\
    A^{(4)}_{2,e}(M_e/M_\tau) &=& 1.83798(33) \cdot 10^{-9}
    \,,\nonumber\\
    A^{(6)}_{2,e}(M_e/M_\tau) &=& - 6.5830(11) \cdot 10^{-8}
    \,,
  \end{eqnarray}
where $M_e/M_\mu=4.83633166(12) \cdot 10^{-3}$
and $M_e/M_\tau=2.87592(26) \cdot 10^{-4}$~\cite{Beringer:1900zz} have been used.
Inserting these values into Eq.~(\ref{eq::A8}) leads to the following
four-loop results
\begin{eqnarray}
  A^{(8)}_{2,e}(M_e/M_\mu) &\approx& (9.161259603 + 0.000711078 + 2.2 \cdot
  10^{-8}) \cdot 10^{-4}\nonumber\\ 
  &\approx& 9.161970703(2)(372) \cdot 10^{-4}
  \,,\nonumber\\
  A^{(8)}_{2,e}(M_e/M_\tau) &\approx& (7.42923268609971 + 2.75209424 \cdot
  10^{-6} + 3.2 \cdot 10^{-13}) \cdot 10^{-6}\nonumber\\
  &\approx& 7.42924(0)(118) \cdot 10^{-6}
  \,,
  \label{eq::Aetaunum}
\end{eqnarray}
where the uncertainty has again been estimated by 10\% of the third term in
the expansion and the parameter uncertainty is displayed separately.
In Ref.~\cite{Aoyama:2012wj} one finds the results\footnote{Note that the
  entry for $A^{(8)}_{2,e}(M_e/M_\tau)$ in Tab.~I of Ref.\cite{Aoyama:2012wj}
  should be multiplied by a factor $1/100$. This misprint has been confirmed by
  the authors of Ref.~\cite{Aoyama:2012wj}.}
$A^{(8)}_{2,e}(M_e/M_\mu)=9.222(66)\times 10^{-4}$ and
$A^{(8)}_{2,e}(M_e/M_\tau)=7.38(12)\times 10^{-6}$  
which agree with our numerical values.

As far as the growth of the coefficients is concerned we observe the same
pattern as for the muon: there is about one order of magnitude between two
and three loops and a factor 100 between three and four loops.
Note, however, that the three-loop result is negative for $a_e$.

\begin{table}[t]
  \begin{center}
    \begin{tabular}{c||l|l}
      group & \multicolumn{2}{c}{$10^4 \cdot A^{(8)}_{2,e}(M_e/M_\mu)$}\\
      \hline
      & this work & \cite{Aoyama:2012wj}\\
      \hline
      I(a) & \hphantom{$-$}0.002264474414(6) & \hphantom{$-$}0.00226456(14)\\
      I(b) + I(c) + II(b) + II(c) & $-1.21390182678(6)$ & $-1.21386(24)$\\
      I(d) & \hphantom{$-$}0.02472687590(2) & \hphantom{$-$}0.024725(7)\\
      III & \hphantom{$-$}8.1715251555(1) & \hphantom{$-$}8.1792(95)\\
      II(a) + IV(d) & $-2.6414355180(7)$ & $-2.642(12)$\\
      IV(a) & \hphantom{$-$}6.3578810372(3) & \hphantom{$-$}6.3583(44)\\
      IV(b) & \hphantom{$-$}0.4157367168(5) & \hphantom{$-$}0.4105(93)\\
      IV(c) & $-1.954826212(2)$ & $-1.897(64)$
    \end{tabular}
  \end{center}
  \caption{\label{tab::e1}Muon mass dependent corrections to
    $a_e$ at four-loop order as obtained in this paper and the 
    comparison with Refs.~\cite{Aoyama:2012wj}. 
    The uncertainties assigned to our numbers correspond to
    10\% of the highest available expansion terms, i.e., the
    ones of order $(M_e/M_\mu)^6$ and $(M_e/M_\mu)^7$.
    Uncertainties from the electron and muon mass are not shown.}
\end{table}

\begin{table}[t]
  \begin{center}
    \begin{tabular}{c||l|l}
      group & \multicolumn{2}{c}{$10^6 \cdot A^{(8)}_{2,e}(M_e/M_\tau)$}\\
      \hline
      & this work & \cite{Aoyama:2012wj}\\
      \hline
      I(a) & \hphantom{$-$}0.0008024665425029(1) & \hphantom{$-$}0.00080233(5)\\
      I(b) + I(c) + II(b) + II(c) & $-0.9458168451136621(8)$ & $-0.94506(25)$\\
      I(d) & \hphantom{$-$}0.0087455060010553(1) & \hphantom{$-$}0.008744(1)\\
      III & \hphantom{$-$}6.059301961911502(2) & \hphantom{$-$}6.061(12)\\
      II(a) + IV(d) & $-1.372489352896281(9)$ & $-1.3835(30)$\\
      IV(a) & \hphantom{$-$}4.510496216222387(2) & \hphantom{$-$}4.5117(69)\\
      IV(b) & \hphantom{$-$}0.147081582099596(4) & \hphantom{$-$}0.1431(95)\\
      IV(c) & $-0.97888609657284(3)$ & $-1.02(11)$
    \end{tabular}
  \end{center}
  \caption{\label{tab::e2}Tau lepton mass dependent corrections to
    $a_e$ at four-loop order as obtained in this paper and the 
    comparison with Refs.~\cite{Aoyama:2012wj}.
    The uncertainties assigned to our numbers correspond to 
    10\% of the highest available expansion terms, i.e., the 
    ones of order $(M_e/M_\tau)^6$ and $(M_e/M_\tau)^7$.
    Uncertainties from the electron and tau lepton mass are not shown.
    The result of Ref.~\cite{Aoyama:2012wj} for the contribution I(d)
    has been multiplied by $1/100$ [see footnote after
    Eq.~(\ref{eq::Aetaunum})].} 
\end{table}

In Tabs.~\ref{tab::e1} and~\ref{tab::e2} our results are shown for the
individual classes of Feynman diagrams. Due to the smallness of the expansion
parameters our method provides an accuracy of at least eight significant
digits. The comparison with the results of
Ref.~\cite{Aoyama:2012wj} demonstrates good overall agreement.
Note that we have applied the methods of
Refs.~\cite{Baikov:1995ui,Baikov:2012rr,Baikov:2013ula}, where four- and
five-loop contributions to $a_\mu$ from polarization function insertions
have been computed, to cross check our result for case I(d).

The quantity $A^{(8)}_{3,e}(M_e/M_\mu,M_e/M_\tau)$
has a more complicated structure since two different heavy
masses are present. However, due to the strong hierarchy
$M_\tau \gg M_\mu \gg M_e$ it is possible to apply the asymptotic expansion
successively which again leads to one-scale vacuum and on-shell integrals.
Our final result reads
\begin{eqnarray}
  A^{(8)}_{3,e}(M_e/M_\mu, M_e/M_\tau) &=&
  \frac{M_e^2}{M_\tau^2} \bigg( -\frac{3123671}{1458000}-\frac{\pi^2}{270}+\frac{\pi^4}{30}-\frac{19 \zeta_3}{45}\nonumber\\&&
  \qquad+\ln\frac{M_\mu^2}{M_\tau^2} \left(\frac{271073}{291600}-\frac{3 \zeta_3}{2}\right)+\frac{89}{810}\ln^2\frac{M_\mu^2}{M_\tau^2} \bigg)\nonumber\\&&
  +\frac{M_e^2 M_\mu}{M_\tau^3} \frac{\pi^2}{90}\nonumber\\&&
  +\frac{M_e^2 M_\mu^2}{M_\tau^4} \bigg(-\frac{1213316893}{5834430000}+\frac{\pi^4}{3150}+\frac{1294 \zeta_3}{3675}-\frac{3}{280}\ln^3\frac{M_\mu^2}{M_\tau^2}\nonumber\\&&
  \qquad+\ln\frac{M_\mu^2}{M_\tau^2} \left(-\frac{9573107}{18522000}+\frac{\zeta_3}{70}\right)+\frac{130813}{1058400}\ln^2\frac{M_\mu^2}{M_\tau^2}\bigg)\nonumber\\&&
  +\frac{M_e^4}{M_\mu^2 M_\tau^2} \bigg( \frac{3304933}{14580000}+\frac{88 \pi^2}{6075}-\frac{107 \zeta_3}{360}+\frac{2533}{40500}\ln\frac{M_\mu^2}{M_\tau^2}\nonumber\\&&
  \qquad+\ln\frac{M_e^2}{M_\mu^2} \left(-\frac{3239}{121500}-\frac{79}{1350}\ln\frac{M_\mu^2}{M_\tau^2}\right)-\frac{7}{8100}\ln^2\frac{M_e^2}{M_\mu^2} \bigg)\nonumber\\&&
  +\frac{M_e^4}{M_\tau^4}\bigg(-\frac{19009349146181}{10081895040000}-\frac{37877173 \zeta_3}{76204800}-\frac{79 \pi^2}{58800}\nonumber\\&&
  \qquad-\frac{373}{40320} P_4+\frac{280111 \pi^4}{14515200}\nonumber\\&&
  \qquad+\ln\frac{M_e^2}{M_\mu^2} \bigg(\frac{441068819}{1714608000}-\frac{33487}{2721600}\ln\frac{M_\mu^2}{M_\tau^2}\nonumber\\&&
  \qquad\qquad\qquad\quad+\frac{1423}{38880}\ln^2\frac{M_\mu^2}{M_\tau^2}-\frac{\pi^2}{420}\bigg)\nonumber\\&&
  \qquad+\ln\frac{M_\mu^2}{M_\tau^2} \left(\frac{767814079}{750141000}-\frac{\pi^2}{420}-\frac{61849 \zeta_3}{80640}\right)\nonumber\\&&
  \qquad-\frac{3034811}{38102400}\ln^2\frac{M_\mu^2}{M_\tau^2}+\frac{1181}{40824}\ln^3\frac{M_\mu^2}{M_\tau^2}\bigg)\nonumber\\&&
  +\frac{M_e^2 M_\mu^3}{M_\tau^5} \frac{\pi^2}{90}+\frac{M_e^4 M_\mu}{M_\tau^5} \left(\frac{79 \pi^2}{19600}+\frac{\pi^2}{140}\ln\frac{M_e^2}{M_\tau^2}\right)
\,.
\end{eqnarray}
After inserting numerical values for the lepton masses one obtains
\begin{eqnarray}
  A^{(8)}_{3,e}(M_e/M_\mu, M_e/M_\tau) &\approx& 
  ( 7.4426 + 0.0261 ) \cdot 10^{-7} \nonumber\\ 
  &\approx& 7.4687(26)(10) \cdot 10^{-7}  \,,
\end{eqnarray}
which has to be compared with
$A^{(8)}_{3,e}(M_e/M_\mu, M_e/M_\tau) = 7.465(18) \cdot 10^{-7}$
as obtained in Ref.~\cite{Aoyama:2012wj}. Again good agreement is found,
however, our analytic result is more precise by about an order of magnitude.

It is interesting to note that the three-loop coefficient which is given
by
\begin{eqnarray}
  A^{(6)}_{3,e}(M_e/M_\mu, M_e/M_\tau) &=& 1.90982(34) \cdot 10^{-13}
  \,,
\end{eqnarray}
is more than six orders of magnitude smaller than the four-loop one which is
due to the fact that the leading term is suppressed by $M_e^4/(M_\mu^2
M_\tau^2)$ whereas at four loops the suppression factor is only
$M_e^2/M_\tau^2$. Note, however, that the overall contribution is very small.

Similarly to the three-loop expression also the leading term of the four-loop
contribution where three one-loop heavy lepton bubbles are inserted into the photon
propagator (see class I(a) in Fig.~\ref{fig::diags})
is of order ${\cal O}(M_e^2/(M_\mu^2
M_\tau^2))$.  Thus we compute for this contribution also the next term of
the hard-mass procedure. It is given by
\begin{eqnarray}
  \lefteqn{\delta A^{(8)}_{3,e}(M_e/M_\mu, M_e/M_\tau)\Bigg|_{I(a),M^6} =}
  \nonumber\\&&
 \frac{M_e^4 M_\mu^2}{M_\tau^6} \bigg(
 -\frac{1032407}{187535250}+\frac{1303}{297675}\ln\frac{M_\mu^2}{M_\tau^2}+\frac{4}{945}\ln^2\frac{M_\mu^2}{M_\tau^2}
\nonumber\\
&&
\qquad\qquad+\ln\frac{M_e^2}{M_\mu^2} \bigg(-\frac{1039}{297675}+\frac{4}{945}\ln\frac{M_\mu^2}{M_\tau^2}\bigg) \bigg)\nonumber\\&&
+\frac{M_e^6}{M_\mu^4 M_\tau^2} \bigg( \frac{204569}{30870000}+\frac{166}{18375}\ln\frac{M_e^2}{M_\mu^2}+\frac{1}{350}\ln^2\frac{M_e^2}{M_\mu^2} \bigg)\nonumber\\&&
+\frac{M_e^6}{M_\mu^2 M_\tau^4} \bigg( \frac{959}{90000}+\frac{31}{5250}\ln\frac{M_e^2}{M_\mu^2}+\frac{1}{350}\ln^2\frac{M_e^2}{M_\mu^2} \bigg)\nonumber\\&&
+\frac{M_e^6}{M_\tau^6} \bigg(\ln\frac{M_e^2}{M_\mu^2} \bigg(\frac{2735573}{187535250}-\frac{199}{297675}\ln\frac{M_\mu^2}{M_\tau^2}+\frac{2}{945}\ln^2\frac{M_\mu^2}{M_\tau^2}\bigg)\nonumber\\&&
\quad\qquad+\frac{8 \zeta_3}{315}-\frac{118286321}{19691201250}+\frac{676036}{31255875}\ln\frac{M_\mu^2}{M_\tau^2}\nonumber\\&&
\quad\qquad+\frac{394}{99225}\ln^2\frac{M_\mu^2}{M_\tau^2}+\frac{2}{945}\ln^3\frac{M_\mu^2}{M_\tau^2} \bigg)
\,.
\end{eqnarray}
This term is included in the numerical values shown in 
Table~\ref{tab::e3} where our results are compared to the ones of
Ref.~\cite{Aoyama:2012wj}. The quality of the agreement is as in the previous
cases.

\begin{table}[t]
  \begin{center}
    \begin{tabular}{c||l|l}
      group & \multicolumn{2}{c}{$10^7 \cdot A^{(8)}_{3,e}(M_e/M_\mu, M_e/M_\tau)$}\\
      \hline
      & this work & ref.\\
      \hline
      I(a) & \hphantom{$-$}0.00001199558(2) & \hphantom{$-$}0.000011994(1)\\
      I(b) + I(c) & \hphantom{$-$}0.172910(24) & \hphantom{$-$}0.172874(21)\\
      II(b) + II(c) & $-1.64747(17)$ & $-1.64866(67)$\\
      IV(a) & \hphantom{$-$}8.9432(25) & \hphantom{$-$}8.941(17)
    \end{tabular}
  \end{center}
  \caption{\label{tab::e3}Lepton mass dependent corrections to
    $a_e$ at four-loop order induced by diagrams which contain
    at the same time the muon and tau lepton.
    The results obtained in this paper are compared to the ones 
    of Refs.~\cite{Aoyama:2012wj}.
    The uncertainties assigned to our numbers correspond to 
    10\% of the highest available expansion terms and
    uncertainties from the lepton masses are not shown.
}
\end{table}
  

\section{\label{sec::concl}Conclusions}

Four-loop corrections induced by a heavy lepton to the anomalous magnetic
moment of the electron and the muon have been computed. This includes tau
lepton contributions to $a_\mu$ and contributions with virtual muons and tau
leptons to $a_e$.  With the help of an asymptotic expansion in the mass ratios
we obtained analytic results. Their numerical evaluation leads to full
agreement with the results of Refs.~\cite{Aoyama:2012wj,Aoyama:2012wk} which
have been obtained with numerical methods. However, our results are more
precise. Actually, the uncertainty is of the order of or even smaller than the
one originating from the imprecise knowledge of the lepton masses. Due to the
decoupling of heavy particles the heavy-lepton contributions are numerically
quite small.



\section*{Acknowledgements}

We would like to thank M.~Nio for useful communications concerning
Ref.~\cite{Aoyama:2012wj}.
This work was supported by the DFG through the SFB/TR~9 ``Computational
Particle Physics'' and by the EU Network
{\sf LHCPHENOnet} PITN-GA-2010-264564. 
The Feynman diagrams were drawn with {\tt
  JaxoDraw}~\cite{Vermaseren:1994je,Binosi:2008ig}.



\begin{appendix}


\section*{\label{app::CTs}Appendix: On-shell counterterms}

In this appendix we provide analytic results for the on-shell counterterms 
for $\alpha$, $M_l$ and $\psi_l$ where the latter stands for the
lepton wave function. We concentrate on the contributions relevant
for our calculation, i.e., the corrections originating from
closed heavy lepton loops.

In the formulae below we use the notation
$L_x = \ln(\mu^2/M_x^2)$ with $x=e,\mu,\tau$ and $a_4=\mbox{Li}_4(1/2)$ 
and mark the contributions from closed electron,
muon and tau loops by the labels $n_e=1$, $n_\mu=1$ and $n_\tau=1$.

In our calculation we renormalize the coupling constant in a first step in
the $\overline{\rm MS}$ scheme and switch to the on-shell scheme after having
obtained a finite result.
The relation between the fine structure constant defined in the 
$\overline{\rm MS}$ scheme, $\bar\alpha(\mu)\equiv \bar\alpha$, and the
corresponding on-shell quantity reads 
\begin{eqnarray}
\frac{\bar{\alpha}}{\alpha} &=& 1
+\frac{\alpha}{\pi} \sum_{i=e,\mu,\tau} \frac{L_i n_i}{3}
+\left(\frac{\alpha}{\pi}\right)^2 \left[\left(\sum_{i=e,\mu,\tau} \frac{L_i n_i}{3}\right)^2 +
  \sum_{i=e,\mu,\tau} \left(\frac{15}{16}+\frac{L_i}{4}\right) n_i\right]
\nonumber\\&&
+\left(\frac{\alpha}{\pi}\right)^3 \Bigg[\left( \sum_{i=e,\mu,\tau} \frac{L_i n_i}{3} \right)^3
+\sum_{\scriptsize \begin{tabular}{c}$i,j=e,\mu,\tau$\\$M_i<M_j$\end{tabular}}
n_i n_j \Bigg(-\frac{311}{1296}-\frac{\pi^2}{18}+L_i
\left(\frac{15}{16}+\frac{5 L_j}{12}\right)
\nonumber\\&&
\qquad\qquad+\frac{23 L_j}{144}+\frac{\pi^2}{6} \left(\frac{M_i}{M_j}\right)+\left(-\frac{167}{150}+\frac{L_i}{45}-\frac{L_j}{45}\right) \left(\frac{M_i}{M_j}\right)^2+\frac{\pi^2}{6} \left(\frac{M_i}{M_j}\right)^3\nonumber\\&&
\qquad\qquad+\bigg(-\frac{23353331}{37044000}-\frac{29 L_i^2}{420}+L_i \left(-\frac{28967}{88200}+\frac{29 L_j}{210}\right)\nonumber\\&&
\qquad\qquad\qquad+\frac{28967 L_j}{88200}-\frac{29 L_j^2}{420}-\frac{\pi^2}{18}\bigg) \left(\frac{M_i}{M_j}\right)^4\nonumber\\&&
\qquad\qquad+\left(\frac{5288963}{62511750}+\frac{2 L_i^2}{315}+L_i \left(\frac{4609}{99225}-\frac{4 L_j}{315}\right)-\frac{4609 L_j}{99225}+\frac{2 L_j^2}{315}\right) \left(\frac{M_i}{M_j}\right)^6\Bigg)\nonumber\\&&
\qquad+\sum_{i=e,\mu,\tau} n_i \left(\frac{77}{576}-\frac{L_i}{32}+\frac{5 \pi^2}{24}-\frac{\ln(2) \pi^2}{3}+\frac{\zeta_3}{192}\right)\nonumber\\&&
\qquad+\sum_{i=e,\mu,\tau} n_i^2 \left(-\frac{695}{648}+\frac{79 L_i}{144}+\frac{5 L_i^2}{24}+\frac{\pi^2}{9}+\frac{7 \zeta_3}{64}\right)\Bigg]
\,,
\end{eqnarray}
where terms up to ${\cal O}(M_i^{8}/M_j^{8})$ are included.
In the case of the heavy lepton contributions to $a_e$ this formula can
immediately be applied, in the case of $a_\mu$ one has to set $n_e=0$.

The bare and on-shell renormalized lepton mass and wave function are
related by
\begin{eqnarray}
  M_l^{\rm bare}    &=& Z_{m,l}^\text{OS} \, M_l \,,\nonumber\\
  \psi_l^{\rm bare} &=& Z_{2,l}^\text{OS} \, \psi_l \,,
\end{eqnarray}
where the renormalization constants for the muon mass and wave function are given by
\begin{eqnarray}
Z_{m,\mu}^\text{OS} &=& 1 + \frac{\bar{\alpha}}{\pi} \Bigg[-1-\frac{3}{4 \epsilon}-\frac{3 L_\mu}{4}+\epsilon \left(-2-L_\mu-\frac{3 L_\mu^2}{8}-\frac{\pi^2}{16}\right)\nonumber\\&&
\qquad+\epsilon^2 \left(-4-\frac{L_\mu^2}{2}-\frac{L_\mu^3}{8}-\frac{\pi^2}{12}+L_\mu \left(-2-\frac{\pi^2}{16}\right)+\frac{\zeta_3}{4}\right)\nonumber\\&&
\qquad+\epsilon^3 \bigg(-8-\frac{L_\mu^3}{6}-\frac{L_\mu^4}{32}-\frac{\pi^2}{6}-\frac{3 \pi^4}{640}+L_\mu^2 \left(-1-\frac{\pi^2}{32}\right)\nonumber\\&&
\qquad\qquad+L_\mu \left(-4-\frac{\pi^2}{12}+\frac{\zeta_3}{4}\right)+\frac{\zeta_3}{3}\bigg)\Bigg]\nonumber\\&&
+\left(\frac{\bar{\alpha}}{\pi}\right)^2 \Bigg[\frac{1}{\epsilon^2} \left(\frac{9}{32}-\frac{n_\mu}{8}-\frac{n_\tau}{8}\right)+\frac{1}{\epsilon} \left(\frac{45}{64}+\frac{9 L_\mu}{16}+\frac{5 n_\mu}{48}+\frac{5 n_\tau}{48}\right)\nonumber\\&&
\qquad+\frac{199}{128}+\frac{45 L_\mu}{32}+\frac{9 L_\mu^2}{16}-\frac{17 \pi^2}{64}+\frac{\ln(2) \pi^2}{2}-\frac{3 \zeta_3}{4}\nonumber\\&&
\qquad+n_\mu \left(\frac{143}{96}+\frac{13 L_\mu}{24}+\frac{L_\mu^2}{8}-\frac{\pi^2}{6}\right)\nonumber\\&&
\qquad+n_\tau \bigg(-\frac{89}{288}+\frac{13 L_\tau}{24}+\frac{L_\mu L_\tau}{4}-\frac{L_\tau^2}{8}+\left(\frac{19}{150}+\frac{L_\mu}{15}-\frac{L_\tau}{15}\right) \left(\frac{M_\mu}{M_\tau}\right)^2\nonumber\\&&
\qquad+\left(\frac{1389}{78400}+\frac{9 L_\mu}{560}-\frac{9 L_\tau}{560}\right) \left(\frac{M_\mu}{M_\tau}\right)^4+\left(\frac{997}{198450}+\frac{2 L_\mu}{315}-\frac{2 L_\tau}{315}\right) \left(\frac{M_\mu}{M_\tau}\right)^6\nonumber\\&&
\qquad+\left(\frac{1229}{627264}+\frac{5 L_\mu}{1584}-\frac{5 L_\tau}{1584}\right) \left(\frac{M_\mu}{M_\tau}\right)^8\bigg)\nonumber\\&&
\qquad+\epsilon \bigg(\frac{677}{256}-12 a_4+\frac{45 L_\mu^2}{32}+\frac{3 L_\mu^3}{8}-\frac{\ln^4(2)}{2}-\frac{205 \pi^2}{128}+3 \ln(2) \pi^2\nonumber\\&&
\quad\qquad-\ln^2(2) \pi^2+\frac{7 \pi^4}{40}-\frac{135 \zeta_3}{16}+L_\mu \left(\frac{199}{64}-\frac{17 \pi^2}{32}+\ln(2) \pi^2-\frac{3 \zeta_3}{2}\right)\nonumber\\&&
\quad\qquad+n_\mu \left(\frac{1133}{192}+\frac{17 L_\mu^2}{24}+\frac{L_\mu^3}{8}-\frac{227 \pi^2}{288}+\ln(2) \pi^2+L_\mu \left(\frac{175}{48}-\frac{5 \pi^2}{16}\right)-\frac{7 \zeta_3}{2}\right)\nonumber\\&&
\quad\qquad+n_\tau \bigg(\frac{869}{1728}+\frac{7 L_\tau}{144}+\frac{L_\mu^2 L_\tau}{8}+\frac{3 L_\tau^2}{8}-\frac{L_\tau^3}{8}+\frac{13 \pi^2}{288}+L_\mu \left(\frac{L_\tau}{3}+\frac{L_\tau^2}{8}+\frac{\pi^2}{48}\right)\nonumber\\&&
\quad\qquad+\left(-\frac{701}{3375}+\frac{L_\mu^2}{30}+L_\mu \left(\frac{1}{30}+\frac{L_\tau}{15}\right)+\frac{11 L_\tau}{50}-\frac{L_\tau^2}{10}\right) \left(\frac{M_\mu}{M_\tau}\right)^2\nonumber\\&&
\quad\qquad+\left(\frac{20481}{10976000}+\frac{9 L_\mu^2}{1120}+L_\mu \left(\frac{27}{1120}+\frac{9 L_\tau}{560}\right)+\frac{111 L_\tau}{9800}-\frac{27 L_\tau^2}{1120}\right) \left(\frac{M_\mu}{M_\tau}\right)^4\nonumber\\&&
\quad\qquad+\left(\frac{584149}{125023500}+\frac{L_\mu^2}{315}+L_\mu \left(\frac{5}{378}+\frac{2 L_\tau}{315}\right)-\frac{631 L_\tau}{198450}-\frac{L_\tau^2}{105}\right) \left(\frac{M_\mu}{M_\tau}\right)^6\nonumber
\\
&&\qquad+\left(\frac{176625767}{60857153280}+\frac{5 L_\mu^2}{3168}+L_\mu \left(\frac{1}{126}+\frac{5 L_\tau}{1584}\right)-\frac{8821 L_\tau}{2195424}-\frac{5 L_\tau^2}{1056}\right) \left(\frac{M_\mu}{M_\tau}\right)^8\bigg)\bigg)\Bigg]\nonumber\\&&
+\left(\frac{\bar{\alpha}}{\pi}\right)^3 \Bigg[\frac{1}{\epsilon^3} \left(-\frac{9}{128}+\frac{3 n_\mu}{32}-\frac{n_\mu^2}{36}-\frac{n_\mu n_\tau}{18}+\frac{3 n_\tau}{32}-\frac{n_\tau^2}{36}\right)\nonumber\\&&
\qquad+\frac{1}{\epsilon^2} \bigg(-\frac{63}{256}-\frac{27 L_\mu}{128}+\left(-\frac{5}{192}+\frac{3 L_\mu}{32}\right) n_\mu+\frac{5 n_\mu^2}{216}\nonumber\\&&
\qquad\qquad+\frac{5 n_\mu n_\tau}{108}+\left(-\frac{5}{192}+\frac{3 L_\mu}{32}\right) n_\tau+\frac{5 n_\tau^2}{216}\bigg)\nonumber\\&&
\qquad+\frac{1}{\epsilon}\Bigg(-\frac{457}{512}-\frac{189 L_\mu}{256}-\frac{81 L_\mu^2}{256}+\frac{35 n_\mu^2}{1296}\nonumber\\&&
\qquad\qquad+\frac{35 n_\mu n_\tau}{648}+\frac{35 n_\tau^2}{1296}+\frac{111 \pi^2}{512}-\frac{3 \ln(2) \pi^2}{8}+\frac{9 \zeta_3}{16}\nonumber\\&&
\qquad\qquad+n_\tau \bigg(\frac{79}{128}+\frac{3 L_\mu^2}{64}+L_\mu \left(\frac{3}{64}-\frac{3 L_\tau}{16}\right)-\frac{13 L_\tau}{32}+\frac{3 L_\tau^2}{32}+\frac{\pi^2}{128}-\frac{\zeta_3}{4}\nonumber\\&&
\qquad\qquad\quad+\left(-\frac{19}{200}-\frac{L_\mu}{20}+\frac{L_\tau}{20}\right) \left(\frac{M_\mu}{M_\tau}\right)^2+\left(-\frac{4167}{313600}-\frac{27 L_\mu}{2240}+\frac{27 L_\tau}{2240}\right) \left(\frac{M_\mu}{M_\tau}\right)^4\nonumber\\&&
\qquad\qquad\quad+\left(-\frac{997}{264600}-\frac{L_\mu}{210}+\frac{L_\tau}{210}\right) \left(\frac{M_\mu}{M_\tau}\right)^6\bigg)\nonumber\\&&
\qquad\qquad+n_\mu \left(-\frac{281}{384}-\frac{23 L_\mu}{64}-\frac{3 L_\mu^2}{64}+\frac{17 \pi^2}{128}-\frac{\zeta_3}{4}\right)\Bigg)\nonumber\\&&
\qquad-\frac{14225}{3072}-3 a_4-\frac{567 L_\mu^2}{512}-\frac{81 L_\mu^3}{256}-\frac{\ln^4(2)}{8}-\frac{6037 \pi^2}{3072}+5 \ln(2) \pi^2+\frac{5 \ln^2(2) \pi^2}{4}\nonumber\\&&
\qquad-\frac{73 \pi^4}{480}+\frac{5 \zeta_5}{8}+\frac{153 \zeta_3}{128}-\frac{\pi^2 \zeta_3}{16}+L_\mu \left(-\frac{1371}{512}+\frac{333 \pi^2}{512}-\frac{9 \ln(2) \pi^2}{8}+\frac{27 \zeta_3}{16}\right)\nonumber\\&&
\qquad+n_\tau \bigg(\frac{6367}{2304}-4 a_4+\frac{L_\mu^3}{64}+L_\mu^2 \left(\frac{3}{128}-\frac{9 L_\tau}{32}\right)-\frac{9 L_\tau^2}{32}+\frac{3 L_\tau^3}{32}\nonumber\\&&
\qquad\qquad-\frac{\ln^4(2)}{6}-\frac{23 \pi^2}{768}+\frac{\ln^2(2) \pi^2}{6}+\frac{11 \pi^4}{360}-\frac{29 \zeta_3}{16}\nonumber\\&&
\qquad\qquad+L_\mu \left(\frac{415}{384}-\frac{21 L_\tau}{32}-\frac{\pi^2}{128}\right)+L_\tau \left(-\frac{1}{64}+\frac{5 \pi^2}{24}-\frac{\ln(2) \pi^2}{3}-\frac{\zeta_3}{4}\right)\nonumber\\&&
\qquad\qquad+\left(\frac{M_\mu}{M_\tau}\right)^2 \bigg(\frac{8153}{12150}-\frac{31 L_\mu^2}{360}+L_\mu \left(-\frac{67}{4050}+\frac{L_\tau}{45}\right)\nonumber\\&&
\qquad\qquad\qquad\qquad\qquad-\frac{4349 L_\tau}{16200}+\frac{23 L_\tau^2}{360}+\frac{2 \pi^2}{135}-\frac{77 \zeta_3}{144}\bigg)\nonumber
\\
&&\qquad+\left(\frac{M_\mu}{M_\tau}\right)^4 \bigg(\frac{13231711}{98784000}-\frac{17 L_\mu^2}{1120}+L_\mu \left(\frac{1907}{44800}-\frac{13 L_\tau}{2240}\right)\nonumber\\&&
\qquad\qquad\qquad\qquad-\frac{517 L_\tau}{6272}+\frac{47 L_\tau^2}{2240}+\frac{\pi^2}{105}-\frac{147 \zeta_3}{1024}\bigg)\nonumber\\&&
\qquad+\left(\frac{M_\mu}{M_\tau}\right)^6 \bigg(\frac{3752184623}{90016920000}-\frac{8 L_\mu^2}{2025}+L_\mu \left(\frac{925261}{35721000}-\frac{181 L_\tau}{28350}\right)\nonumber\\&&
\qquad\qquad\qquad\qquad-\frac{664523 L_\tau}{17860500}+\frac{293 L_\tau^2}{28350}+\frac{32 \pi^2}{6075}-\frac{119 \zeta_3}{1920}\bigg)\bigg)\nonumber\\&&
+n_\tau^2 \bigg(\frac{1685}{7776}+\frac{31 L_\tau}{108}-\frac{13 L_\tau^2}{72}-\frac{L_\mu L_\tau^2}{12}+\frac{L_\tau^3}{18}-\frac{7 \zeta_3}{18}\nonumber\\&&
\qquad+\left(\frac{23}{324}-\frac{19 L_\tau}{225}-\frac{2 L_\mu L_\tau}{45}+\frac{2 L_\tau^2}{45}\right) \left(\frac{M_\mu}{M_\tau}\right)^2\nonumber\\&&
\qquad+\left(-\frac{119}{24000}+L_\mu \left(-\frac{1}{100}-\frac{3 L_\tau}{280}\right)-\frac{71 L_\tau}{39200}+\frac{3 L_\tau^2}{280}\right) \left(\frac{M_\mu}{M_\tau}\right)^4\nonumber\\&&
\qquad+\left(-\frac{1594}{496125}+L_\mu \left(-\frac{1}{175}-\frac{4 L_\tau}{945}\right)+\frac{704 L_\tau}{297675}+\frac{4 L_\tau^2}{945}\right) \left(\frac{M_\mu}{M_\tau}\right)^6\bigg)\nonumber\\&&
+n_\mu n_\tau \bigg(-\frac{1327}{3888}-\frac{13 L_\mu L_\tau}{36}-\frac{L_\mu^2 L_\tau}{12}+\frac{L_\tau^3}{36}+L_\tau \left(-\frac{5}{8}+\frac{\pi^2}{9}\right)+\frac{2 \zeta_3}{9}\nonumber\\&&
\qquad+\left(-\frac{1541}{3375}-\frac{11 L_\mu}{45}-\frac{L_\mu^2}{45}+\frac{4 L_\tau}{25}+\frac{L_\tau^2}{45}+\frac{4 \pi^2}{135}\right) \left(\frac{M_\mu}{M_\tau}\right)^2\nonumber\\&&
\qquad+\left(-\frac{1833259}{16464000}-\frac{9 L_\mu^2}{560}+L_\mu \left(-\frac{3551}{39200}+\frac{3 L_\tau}{140}\right)+\frac{193 L_\tau}{2450}-\frac{3 L_\tau^2}{560}\right) \left(\frac{M_\mu}{M_\tau}\right)^4\nonumber\\&&
\qquad+\left(-\frac{1997398}{93767625}-\frac{2 L_\mu^2}{315}+L_\mu \left(-\frac{8021}{297675}+\frac{8 L_\tau}{945}\right)+\frac{7024 L_\tau}{297675}-\frac{2 L_\tau^2}{945}\right) \left(\frac{M_\mu}{M_\tau}\right)^6\bigg)\nonumber\\&&
+n_\mu \bigg(-\frac{5257}{2304}+\frac{8 a_4}{3}-\frac{117 L_\mu^2}{128}-\frac{11 L_\mu^3}{64}+\frac{\ln^4(2)}{9}-\frac{1327 \pi^2}{6912}+\frac{37 \zeta_3}{96}\nonumber\\&&
\qquad+\frac{5 \ln(2) \pi^2}{36}-\frac{\ln^2(2) \pi^2}{9}+\frac{91 \pi^4}{2160}+L_\mu \left(-\frac{1145}{384}+\frac{221 \pi^2}{384}-\frac{\ln(2) \pi^2}{3}-\frac{\zeta_3}{4}\right)\bigg)\nonumber\\&&
+n_\mu^2 \left(-\frac{9481}{7776}-\frac{13 L_\mu^2}{72}-\frac{L_\mu^3}{36}+\frac{4 \pi^2}{135}+L_\mu \left(-\frac{197}{216}+\frac{\pi^2}{9}\right)+\frac{11 \zeta_3}{18}\right)\Bigg]
\,,
\end{eqnarray}

\begin{eqnarray}
Z_{2,\mu}^\text{OS} &=& 1 + \frac{\bar{\alpha}}{\pi} \Bigg[-1-\frac{3}{4 \epsilon}-\frac{3 L_\mu}{4}+\epsilon \left(-2-L_\mu-\frac{3 L_\mu^2}{8}-\frac{\pi^2}{16}\right)\nonumber\\&&
\qquad+\epsilon^2 \left(-4-\frac{L_\mu^2}{2}-\frac{L_\mu^3}{8}-\frac{\pi^2}{12}+L_\mu \left(-2-\frac{\pi^2}{16}\right)+\frac{\zeta_3}{4}\right)\nonumber\\&&
\qquad+\epsilon^3 \bigg(-8-\frac{L_\mu^3}{6}-\frac{L_\mu^4}{32}-\frac{\pi^2}{6}-\frac{3 \pi^4}{640}+L_\mu^2 \left(-1-\frac{\pi^2}{32}\right)\nonumber\\&&
\qquad\qquad+L_\mu \left(-4-\frac{\pi^2}{12}+\frac{\zeta_3}{4}\right)+\frac{\zeta_3}{3}\bigg)\Bigg]\nonumber\\&&
+\left(\frac{\bar{\alpha}}{\pi}\right)^2 \Bigg[\frac{9}{32 \epsilon^2}+\frac{1}{\epsilon} \left(\frac{51}{64}+\frac{9 L_\mu}{16}+\left(\frac{1}{16}+\frac{L_\mu}{4}\right) n_\mu+\left(\frac{1}{16}+\frac{L_\tau}{4}\right) n_\tau\right)\nonumber\\&&
\qquad+\frac{433}{128}+\frac{51 L_\mu}{32}+\frac{9 L_\mu^2}{16}-\frac{49 \pi^2}{64}+\ln(2) \pi^2-\frac{3 \zeta_3}{2}\nonumber\\&&
\qquad+n_\mu \left(\frac{947}{288}+\frac{11 L_\mu}{24}+\frac{3 L_\mu^2}{8}-\frac{5 \pi^2}{16}\right)\nonumber\\&&
\qquad+n_\tau \bigg(\frac{\pi^2}{48}-\frac{5}{96}+\frac{11 L_\tau}{24}+\frac{L_\mu L_\tau}{4}+\frac{L_\tau^2}{8}+\frac{1}{15}\left(\frac{M_\mu}{M_\tau}\right)^2\nonumber\\&&
\qquad\qquad+\left(-\frac{129}{78400}-\frac{9 L_\mu}{560}+\frac{9 L_\tau}{560}\right) \left(\frac{M_\mu}{M_\tau}\right)^4\nonumber\\&&
\qquad\qquad+\left(-\frac{367}{99225}-\frac{4 L_\mu}{315}+\frac{4 L_\tau}{315}\right) \left(\frac{M_\mu}{M_\tau}\right)^6\nonumber\\&&
\qquad\qquad+\left(-\frac{569}{209088}-\frac{5 L_\mu}{528}+\frac{5 L_\tau}{528}\right) \left(\frac{M_\mu}{M_\tau}\right)^8\bigg)\nonumber\\&&
\qquad+\epsilon \bigg(\frac{211}{256}+\frac{51 L_\mu^2}{32}+\frac{3 L_\mu^3}{8}-\ln^4(2)-\frac{339 \pi^2}{128}+\frac{23 \ln(2) \pi^2}{4}-2 \ln^2(2) \pi^2\nonumber\\&&
\qquad\qquad+L_\mu \left(\frac{433}{64}-\frac{49 \pi^2}{32}+2 \ln(2) \pi^2-3 \zeta_3\right)-\frac{297 \zeta_3}{16}+\frac{7 \pi^4}{20}-24 a_4\nonumber\\&&
\qquad\qquad+n_\mu \bigg(\frac{17971}{1728}+\frac{5 L_\mu^2}{8}+\frac{7 L_\mu^3}{24}-\frac{445 \pi^2}{288}+2 \ln(2) \pi^2\nonumber\\&&
\qquad\qquad\qquad\quad+L_\mu \left(\frac{1043}{144}-\frac{29 \pi^2}{48}\right)-\frac{85 \zeta_3}{12}\bigg)\nonumber\\&&
\qquad\qquad+n_\tau \bigg(\frac{89}{576}+\frac{L_\mu^2 L_\tau}{8}+\frac{7 L_\tau^2}{24}+\frac{L_\tau^3}{24}+\frac{11 \pi^2}{288}-\frac{\zeta_3}{12}+L_\tau \left(\frac{9}{16}+\frac{\pi^2}{24}\right)\nonumber
\\
&&\qquad+L_\mu \left(\frac{L_\tau}{3}+\frac{L_\tau^2}{8}+\frac{\pi^2}{48}\right)+\left(-\frac{7}{75}+\frac{2 L_\tau}{15}\right) \left(\frac{M_\mu}{M_\tau}\right)^2\nonumber\\&&
\qquad+\left(\frac{49659}{10976000}-\frac{9 L_\mu^2}{1120}+L_\mu \left(-\frac{27}{1120}-\frac{9 L_\tau}{560}\right)+\frac{51 L_\tau}{2450}+\frac{27 L_\tau^2}{1120}\right) \left(\frac{M_\mu}{M_\tau}\right)^4\nonumber\\&&
\qquad+\left(-\frac{71329}{62511750}-\frac{2 L_\mu^2}{315}+L_\mu \left(-\frac{5}{189}-\frac{4 L_\tau}{315}\right)+\frac{1891 L_\tau}{99225}+\frac{2 L_\tau^2}{105}\right) \left(\frac{M_\mu}{M_\tau}\right)^6\nonumber\\&&
\qquad+\bigg(-\frac{55373867}{20285717760}-\frac{5 L_\mu^2}{1056}+L_\mu \left(-\frac{1}{42}-\frac{5 L_\tau}{528}\right)\nonumber\\&&
\qquad\qquad+\frac{13441 L_\tau}{731808}+\frac{5 L_\tau^2}{352}\bigg) \left(\frac{M_\mu}{M_\tau}\right)^8\bigg)\bigg)\Bigg]\nonumber\\&&
+\left(\frac{\bar{\alpha}}{\pi}\right)^3 \Bigg[-\frac{9}{128 \epsilon^3}+\frac{1}{\epsilon^2} \bigg(-\frac{81}{256}-\frac{27 L_\mu}{128}+\left(-\frac{7}{192}-\frac{3 L_\mu}{16}\right) n_\mu+\frac{n_\mu^2}{72}\nonumber\\&&
\qquad\qquad\qquad\qquad\qquad+\frac{n_\mu n_\tau}{36}+\left(-\frac{7}{192}-\frac{3 L_\tau}{16}\right) n_\tau+\frac{n_\tau^2}{72}\bigg)\nonumber\\&&
\qquad+\frac{1}{\epsilon}\bigg(-\frac{1039}{512}-\frac{243 L_\mu}{256}-\frac{81 L_\mu^2}{256}+\frac{303 \pi^2}{512}-\frac{3 \ln(2) \pi^2}{4}+\frac{9 \zeta_3}{8}\nonumber\\&&
\qquad\qquad+\left(-\frac{5}{432}-\frac{L_\mu^2}{12}\right) n_\mu^2+\left(-\frac{5}{216}-\frac{L_\mu L_\tau}{6}\right) n_\mu n_\tau+\left(-\frac{5}{432}-\frac{L_\tau^2}{12}\right) n_\tau^2\nonumber\\&&
\qquad\qquad+n_\tau \bigg(\frac{85}{128}+L_\mu \left(-\frac{3}{64}-\frac{3 L_\tau}{8}\right)-\frac{13 L_\tau}{32}-\frac{3 L_\tau^2}{32}-\frac{\pi^2}{64}\nonumber\\&&
\qquad\qquad\qquad-\frac{1}{20} \left(\frac{M_\mu}{M_\tau}\right)^2+\left(\frac{387}{313600}+\frac{27 L_\mu}{2240}-\frac{27 L_\tau}{2240}\right) \left(\frac{M_\mu}{M_\tau}\right)^4\nonumber\\&&
\qquad\qquad\qquad+\left(\frac{367}{132300}+\frac{L_\mu}{105}-\frac{L_\tau}{105}\right) \left(\frac{M_\mu}{M_\tau}\right)^6\bigg)\nonumber\\&&
\qquad\qquad+n_\mu \left(-\frac{707}{384}-\frac{29 L_\mu}{64}-\frac{15 L_\mu^2}{32}+\frac{15 \pi^2}{64}\right)\bigg)\nonumber\\&&
\qquad-\frac{10823}{3072}-\frac{729 L_\mu^2}{512}-\frac{81 L_\mu^3}{256}-\frac{5 \ln^4(2)}{12}-\frac{58321 \pi^2}{9216}-10 a_4\nonumber\\&&
\qquad+\frac{685 \ln(2) \pi^2}{48}+3 \ln^2(2) \pi^2-\frac{41 \pi^4}{120}-\frac{739 \zeta_3}{128}+\frac{\pi^2 \zeta_3}{8}\nonumber\\&&
\qquad+L_\mu \left(-\frac{3117}{512}+\frac{909 \pi^2}{512}-\frac{9 \ln(2) \pi^2}{4}+\frac{27 \zeta_3}{8}\right)-\frac{5 \zeta_5}{16}\nonumber
\\
&&+n_\tau^2 \bigg(-\frac{35}{2592}-\frac{11 L_\tau^2}{72}-\frac{L_\mu L_\tau^2}{12}-\frac{L_\tau^3}{12}-\frac{L_\tau \pi^2}{72}-\frac{2 L_\tau}{45} \left(\frac{M_\mu}{M_\tau}\right)^2\nonumber\\&&
\qquad+\left(-\frac{121}{24000}+L_\mu \left(\frac{1}{100}+\frac{3 L_\tau}{280}\right)-\frac{349 L_\tau}{39200}-\frac{3 L_\tau^2}{280}\right) \left(\frac{M_\mu}{M_\tau}\right)^4\nonumber\\&&
\qquad+\left(\frac{353}{496125}+L_\mu \left(\frac{2}{175}+\frac{8 L_\tau}{945}\right)-\frac{2668 L_\tau}{297675}-\frac{8 L_\tau^2}{945}\right) \left(\frac{M_\mu}{M_\tau}\right)^6\bigg)\nonumber\\&&
+n_\mu n_\tau \bigg(-\frac{35}{1296}-\frac{L_\mu^2 L_\tau}{4}+L_\tau \left(-\frac{481}{216}+\frac{5 \pi^2}{24}\right)\nonumber\\&&
\qquad+L_\mu \left(-\frac{11 L_\tau}{36}-\frac{L_\tau^2}{12}-\frac{\pi^2}{72}\right)+\left(-\frac{37}{90}-\frac{2 L_\mu}{45}+\frac{2 \pi^2}{45}\right) \left(\frac{M_\mu}{M_\tau}\right)^2\nonumber\\&&
\qquad+\left(-\frac{317689}{16464000}-\frac{3 L_\mu}{1120}+\frac{3 L_\mu^2}{560}+\frac{37 L_\tau}{9800}-\frac{3 L_\tau^2}{560}\right) \left(\frac{M_\mu}{M_\tau}\right)^4\nonumber\\&&
\qquad+\left(-\frac{159337}{18753525}+\frac{151 L_\mu}{14175}+\frac{4 L_\mu^2}{945}-\frac{2437 L_\tau}{297675}-\frac{4 L_\tau^2}{945}\right) \left(\frac{M_\mu}{M_\tau}\right)^6\bigg)\nonumber\\&&
+n_\mu^2 \left(-\frac{8425}{2592}-\frac{11 L_\mu^2}{72}-\frac{L_\mu^3}{6}+\frac{2 \pi^2}{45}+L_\mu \left(-\frac{481}{216}+\frac{5 \pi^2}{24}\right)+\frac{7 \zeta_3}{3}\right)\nonumber\\&&
+n_\tau \bigg(\frac{4285}{2304}+L_\mu^2 \left(-\frac{3}{128}-\frac{3 L_\tau}{8}\right)-\frac{L_\tau^2}{8}-\frac{L_\tau^3}{32}-\frac{29 \pi^2}{768}+\frac{3 \zeta_3}{16}\nonumber\\&&
\qquad+L_\mu \left(\frac{87}{128}-\frac{31 L_\tau}{32}-\frac{3 L_\tau^2}{16}-\frac{\pi^2}{32}\right)+L_\tau \left(-\frac{133}{192}+\frac{95 \pi^2}{192}-\frac{2 \ln(2) \pi^2}{3}+\zeta_3\right)\nonumber\\&&
\qquad+\left(\frac{M_\mu}{M_\tau}\right)^2 \left(-\frac{461}{1296}-\frac{71 L_\mu}{360}+\frac{17 L_\tau}{360}+\frac{\pi^2}{45}\right)\nonumber\\&&
\qquad+\left(\frac{M_\mu}{M_\tau}\right)^4 \bigg(-\frac{2229623}{8232000}+\frac{3 L_\mu^2}{80}+L_\mu \left(-\frac{28607}{940800}-\frac{87 L_\tau}{2240}\right)\nonumber\\&&
\qquad\qquad\qquad\qquad+\frac{3209 L_\tau}{94080}+\frac{3 L_\tau^2}{2240}+\frac{\pi^2}{105}+\frac{147 \zeta_3}{1024}\bigg)\nonumber\\&&
\qquad+\left(\frac{M_\mu}{M_\tau}\right)^6 \bigg(-\frac{15665719421}{90016920000}+\frac{1913 L_\mu^2}{56700}+L_\mu \left(-\frac{1587179}{71442000}-\frac{1103 L_\tau}{28350}\right)\nonumber\\&&
\qquad\qquad\qquad\qquad+\frac{2181719 L_\tau}{71442000}+\frac{293 L_\tau^2}{56700}+\frac{16 \pi^2}{6075}+\frac{119 \zeta_3}{960}\bigg)\bigg)\nonumber\\&&
+n_\mu \bigg(-\frac{76897}{6912}+12 a_4-\frac{143 L_\mu^2}{128}-\frac{19 L_\mu^3}{32}+\frac{\ln^4(2)}{2}-\frac{11551 \pi^2}{20736}+\frac{7 \ln(2) \pi^2}{18}\nonumber\\&&
\qquad-\frac{\ln^2(2) \pi^2}{2}+\frac{31 \pi^4}{720}+\frac{1763 \zeta_3}{288}+L_\mu \left(-\frac{2891}{384}+\frac{233 \pi^2}{192}-\frac{2 \ln(2) \pi^2}{3}+\zeta_3\right)\bigg)\Bigg]
\,,
\nonumber\\
\end{eqnarray}
{where we include terms up to ${\cal O}(1/M_\tau^{8})$.}

The mass and wave function renormalization constants for the electron can be
constructed from the above results by replacing $M_\mu$ by $M_e$, $L_\mu$ by
$L_e$ and $n_\mu$ by $n_e$. 
Moreover the terms proportional to $n_\tau$ and $n_\tau^2$ have to be
duplicated and afterwards the replacements $n_\tau\to n_\mu$, $M_\tau \to
M_\mu$ and $L_\tau \to L_\mu$ have to be performed 
in one of the expressions. Furthermore, one has to
add the contributions involving simultaneously virtual muon and tau loops
which are given by
\begin{eqnarray}
\delta Z_{m,e}^\text{OS} &=&\left(\frac{\bar{\alpha}}{\pi}\right)^3 \bigg[-\frac{1}{18 \epsilon^3}+\frac{5}{108 \epsilon^2}+\frac{35}{648 \epsilon}-\frac{1327}{3888}+\frac{2 \zeta_3}{9}\nonumber\\&&
\quad+\frac{31 L_\tau}{54}-\frac{13 L_\mu L_\tau}{36}-\frac{L_e L_\mu L_\tau}{6}+\frac{L_\mu^2 L_\tau}{12}+\frac{L_\tau^3}{36}\nonumber\\&&
\quad+\frac{M_\mu^2}{M_\tau^2} \left( -\frac{47}{150}-\frac{L_\mu}{5}+\frac{L_\tau}{5} \right)+\frac{M_e^2}{M_\mu^2} \left( -\frac{19 L_\tau}{225}-\frac{2 L_e L_\tau}{45}+\frac{2 L_\mu L_\tau}{45} \right)\nonumber\\&&
\quad+\frac{M_e^2}{M_\tau^2} \left( \frac{1937}{10125}-\frac{2 L_\mu}{45}-\frac{2 L_e L_\mu}{45}+\frac{L_\mu^2}{45}-\frac{L_\tau}{25}+\frac{L_\tau^2}{45} \right)\nonumber\\&&
\quad+\frac{M_e^2 M_\mu^2}{M_\tau^4} \left( -\frac{107}{3675}-\frac{L_\mu}{35}+\frac{L_\tau}{35} \right)+\frac{M_e^4}{M_\mu^2 M_\tau^2} \left( -\frac{33}{1000}-\frac{L_e}{50}+\frac{L_\mu}{50} \right)\nonumber\\&&
\quad+\frac{M_\mu^4}{M_\tau^4} \left(-\frac{224261}{2058000}-\frac{529 L_\mu}{9800}-\frac{3 L_\mu^2}{280}+\frac{529 L_\tau}{9800}+\frac{3 L_\mu L_\tau}{140}-\frac{3 L_\tau^2}{280}\right)\nonumber\\&&
\quad+\frac{M_e^4}{M_\mu^4} \left( -\frac{463 L_\tau}{39200}-\frac{3 L_e L_\tau}{280}+\frac{3 L_\mu L_\tau}{280} \right)\nonumber\\&&
\quad+\frac{M_e^4}{M_\tau^4} \left( \frac{39379}{1029000}-\frac{9 L_\mu}{1120}-\frac{3 L_e L_\mu}{280}+\frac{3 L_\mu^2}{560}-\frac{37 L_\tau}{9800}+\frac{3 L_\tau^2}{560} \right)
\bigg]
\,,
\end{eqnarray}
\begin{eqnarray}
\delta Z_{2,e}^\text{OS} &=&\left(\frac{\bar{\alpha}}{\pi}\right)^3 \bigg[ \frac{1}{36 \epsilon^2}+\frac{1}{\epsilon} \left(-\frac{5}{216}-\frac{L_\mu L_\tau}{6}\right)-\frac{35}{1296}-\frac{11 L_\mu L_\tau}{36}\nonumber\\&&
\quad-\frac{L_e L_\mu L_\tau}{6}-\frac{L_\mu^2 L_\tau}{12}-\frac{L_\mu L_\tau^2}{12}-\frac{L_\mu \pi^2}{72}-\frac{L_\tau \pi^2}{72}\nonumber\\&&
\quad-\frac{2 L_\tau}{45} \frac{M_e^2}{M_\mu^2}-\frac{2 L_\mu}{45} \frac{M_e^2}{M_\tau^2}+\left( \frac{13}{1000}+\frac{L_e}{50}-\frac{L_\mu}{50} \right)\frac{M_e^4}{M_\mu^2 M_\tau^2}\nonumber\\&&
\quad+\frac{M_e^4}{M_\mu^4} \left( \frac{43 L_\tau}{39200}+\frac{3 L_e L_\tau}{280}-\frac{3 L_\mu L_\tau}{280} \right)\nonumber\\&&
\quad+\frac{M_e^4}{M_\tau^4} \left( -\frac{39379}{1029000}-\frac{3 L_\mu}{1120}+\frac{3 L_e L_\mu}{280}-\frac{3 L_\mu^2}{560}+\frac{37 L_\tau}{9800}-\frac{3 L_\tau^2}{560} \right)
\bigg]
\,.
\end{eqnarray}
These formulae include only terms up to quartic order in the inverse heavy mass since
the corresponding contributions to $a_e$ are only computed up to this order.


\end{appendix}




\end{document}